\documentclass{osa-article}

\usepackage{bm}
\usepackage{graphicx}
\usepackage{array}
\usepackage{dcolumn}
\usepackage{amsmath}
\usepackage{esint}

\usepackage{multirow}
\usepackage{hyperref}
\usepackage{mathrsfs}
\usepackage{makecell}
\usepackage{threeparttable}
\usepackage{graphicx}
\newcommand{\beq}{\begin{equation}}
\newcommand{\beql}[1]{\begin{equation}\label{#1}}

\newcommand{\eeq}{\end{equation}}
\newcommand{\bsp}{\begin{split}}
\newcommand{\esp}{\end{split}}
\newcommand{\Eq}[1]{Eq.~(\ref{#1})}

\newcommand{\Fig}[1]{Fig.~\ref{#1}}
\newcommand{\Figure}[1]{Figure~\ref{#1}}

\usepackage{todonotes}

\journal{oe}

\begin{document}

\title{Numerical approximation of slowlingly varying envelope in finite element electromagnetism: a ray-wave method of modeling multi-scale devices }

\author{Fan Xiao,\authormark{1} Jingwei Wang, \authormark{1}  Zhongfei Xiong, \authormark{1} and Yuntian Chen\authormark{1,2,*}}

\address{\authormark{1}School of Optical and Electronic Information, Huazhong University of Science and Technology, Wuhan 430074, China\\
\authormark{2}Wuhan National Laboratory of Optoelectronics, Huazhong University of Science and Technology, Wuhan 430074, China\\}

\email{\authormark{*} yuntian@hust.edu.cn} 


\begin{abstract}
In this work we propose an efficient and accurate multi-scale optical simulation algorithm by applying a numerical version of slowly varying envelope approximation in FEM. Specifically, we employ the fast iterative method to quickly compute the phase distribution of the electric field within the computational domain and construct a novel multi-scale basis function that combines the conventional polynomial basis function  together with numerically resolved phase information of optical waves. Utilizing this multi-scale basis function, the finite element method can significantly reduce the degrees of freedom required for the solution while maintaining computational accuracy, thereby improving computational efficiency. Without loss of generality, we illustrate our approach via simulating the examples of lens groups and gradient-index lenses, accompanied with performance benchmark against the standard finite element method. The results demonstrate that the proposed method achieves consistent results with the standard finite element method but with a computational speed improved by an order of magnitude.
\end{abstract}


\section{Introduction}
\begin{figure}[h]
    \centering
    \includegraphics[width=1\linewidth]{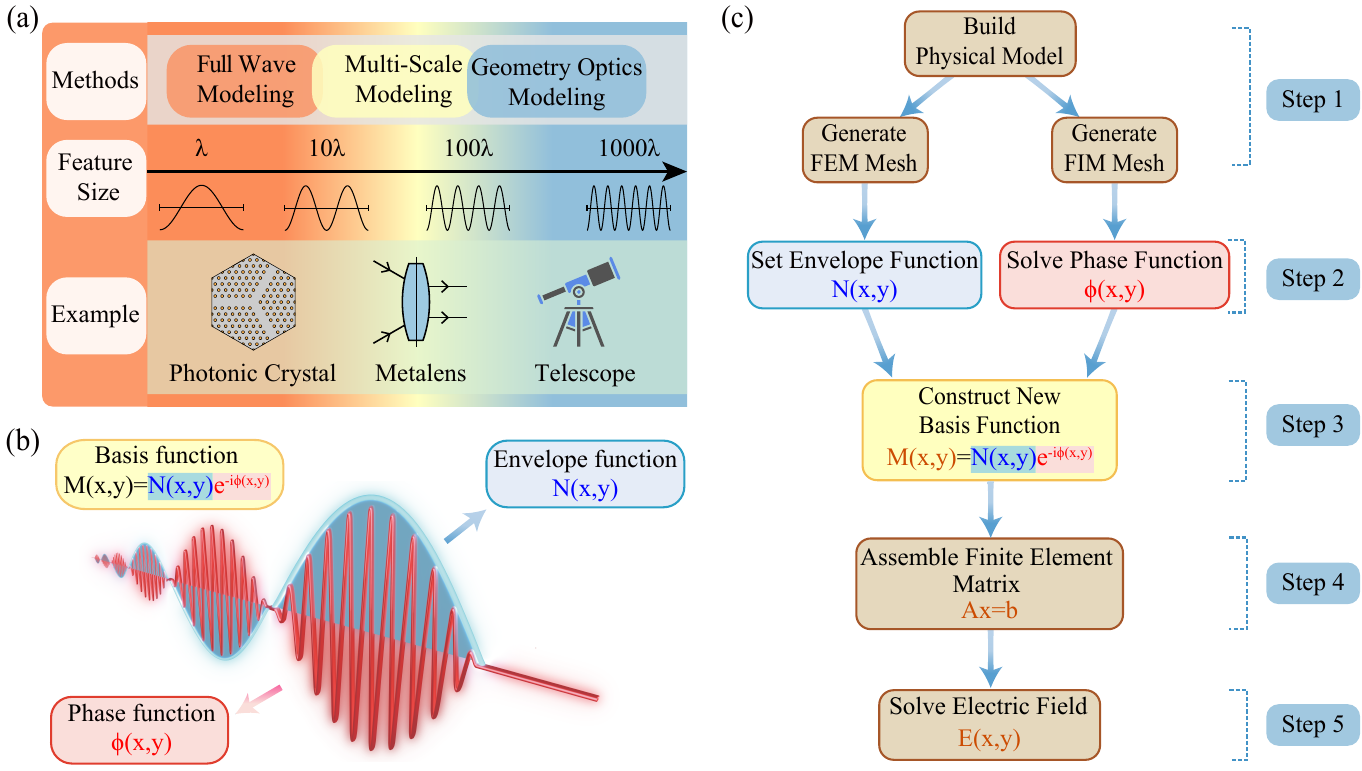}
    \caption{(a) Length scales, device examples of fullwave/multi-scale/geometry optics modeling, (b) basic idea of double-wave method for multi-scale devices, (c) workflow of RWM algorithm.}
    \label{fig1}
\end{figure}

In the modeling of photonic devices, fullwave modeling techniques, i.e., Finite Difference Time Domain (FDTD)\cite{taflove2005computational} or Finite Element Method (FEM)\cite{jin2015finite}, usually solves the vector wave equation with typical computational domain ranging from a few wavelengths to several hundred by maximum\cite{kang2024large}, while geometry optics modeling can model large-sized optical devices/systems beyond thousand wavelengths, as sketch in \Fig{fig1}(a). Either the fullwave modeling or geometry optics modeling via Ray-Tracing (RT) technique\cite{glassner1989introduction}, focuses on limited length scale of photonic devices, and becomes inadequate to model complex photonic devices/systems, especially those with multi-scale feature sizes. However, the recent trend of photonic devices evolves towards two distinct research directions:  1) photonic devices underpinned by novel mechanism that contains complex and effective medium; 2) functional devices that continuously demand miniaturization and integration. For instance, metalenses\cite{pan2022dielectric,wang2021high,yoon2020single}, i.e, seen as an effective yet thin medium, control the flow of light via surfaces and can be flat and thin, thus promise to scale down the size of the optical systems, especially in the context of technology and commercialization  development of Augmented/Virtual Reality (AR/VR)\cite{ding2023waveguide,xiong2021augmented,zou2021doubling,rolland2024waveguide}. Nevertheless, metalens development is challenged by the complexity of the design workflow, which usually consists of simulating  millions of subwavelength unit-cells called meta-atoms, in junction with millimeter-scale surface hybrid with traditional refractive lens. Another relevant example is Light Detection and Ranging (LiDAR)\cite{li2022progress,kim2021nanophotonics}, which also integrates photonic integrated circuits with traditional optical components. Evidently, both of the two trends have crying needs for multi-scale modeling for navigating the intricacies of photonic engineering that combines advantages of numerical accuracy of fullwave modeling and the large scale capability of geometry optics modeling.

The current mainstream approach for multi-scale optical systems is to use hybrid methods that combine geometrical optics algorithms with full-wave algorithms for simulation. For instance, Leiner et al. integrated RT with FDTD, each simulating different scale regions and exchanging data through predefined  files\cite{leiner2013multi,leiner2012simulation,leiner2014simulation,leiner2014multiple}. Similarly, companies like Ansys and Synoposys\cite{ansys} adopt this approach by employing their respective geometrical optics software, like Ansys Zemax OpticStudio or CODE V, alongside wave optics software, such as Ansys Lumerical FDTD or RSoft Photonic Device Tools, to simulate devices of different sizes, achieving multi-scale optical system simulation. Additionally, Ziga Lokar et al. employed a combination of Rigorous Coupled Wave Analysis (RCWA), RT, and transfer matrix method to simulate solar cells\cite{lokar2018performance,lokar2019coupled}. Furthermore, Frank Wyrowski et al. combined the Fourier transform with RT to handle the electric field at micro-nano structures\cite{wyrowski2011introduction,kuhn2013non}. Despite the advantages of hybrid methods in addressing multi-scale problems by integrating different algorithms, challenges remain. These include the difficulties of dealing with multiple light sources and ensuring mathematical and physical consistency when exchanging electromagnetic field information between different algorithms.

In this article, we attempt to solve photonic multi-scale problems by encoding the wavefront propagation information solved directly from eikonal equations into the finite element algorithms inherently. Since solving eikonal equations is equivalent to geometry optics modeling via RT, thus our proposed approach, coined as the ray-wave method (RWM) onwards, essentially combines advantages from both fullwave and geometry optics modeling. Here, \textit{ray-wave} refers to encode the wavefront information calculated from optical path of \textit{ray} bundles into \textit{fullwave} modeling algorithms. The ray-wave method can be seen as a numerical version of slowly varying envelope approximation (SVEA) in FEM\cite{hoekstra1993new,wang2024multi,morimoto2021efficient}. In contrast to SVEA-FEM, wherein the phase information is analytically given if the light beam propagates along a main direction, the phase information in RWM is obtained by  numerically solve the eikonal equation using Fast Iterative Method (FIM)\cite{jeong2008fast,fu2011fast,fu2013fast}, thereby RWM has no restriction on one major directional or bidirectional  propagation as assumed in SVEA-FEM. As such, our RWM is capable of achieving the high numerical accuracy comparable to fullwave modeling and sizable computational domain guaranteed by FIM simultaneously for considerably large and multi-scale photonic devices.

The paper is organized as follows, in Section 2 we present the principles of RWM as well as the processes of the algorithm in detail. In Section 3 we apply RWM to simulate several examples and compare the results with the standard FEM. Finally, Section 4 concludes the paper.

\section{Theory and methods}
\subsection{General Framework}
All the simulations associated with photonics engineering can be boiled to solve the the following  vector wave equation approximated at different levels of simplicity,
\begin{align}\label{wave_eq}
\nabla\times \bar{\bm {\mu}}_r^{-1} (\bm r) \nabla\times \boldsymbol{E}(\boldsymbol{r})-k_{0}^2 \bar{\bm {\epsilon}}_r (\bm r) \boldsymbol{E}(\boldsymbol{r})=0,
\end{align}
where $\boldsymbol{E}(\boldsymbol{r})$ is the electric field, $\bar{\bm {\epsilon}}_r (\bm r)$/$\bar{\bm {\mu}}_r(\bm r)$ represents the relative permittivity/permeability of the medium. Here, $k_0=2\pi/\lambda_0$, where $k_0$ and $\lambda_0$ represent the wave number and wavelength in free space, respectively. Without any approximation, the original vector wave equation given in \Eq{wave_eq} is most difficult to solve. One of extremely successful approximation is the geometry optics approximation, wherein \Eq{wave_eq} is reduced to scale wave equation  and is further approximated as the eikonal equation under the assumption that the field amplitude stays unchanged locally. Notably, the eikonal equation essentially extracts the phase information of propagating waves with incredible accuracy, especially in  large scale scenario wherein the optical inhomogeneity is weak.  Inspired by this simple observation, we decompose the electric field $\boldsymbol{E}(\boldsymbol{r})$  into two components sketched in \Fig{fig1}(b), i.e., $\boldsymbol{E}(\boldsymbol{r})  =\boldsymbol{e}(\boldsymbol{r})e^{-i\phi(\boldsymbol{r})}$, which contains a rapidly varying phase factor $e^{-i\phi(\boldsymbol{r})}$  and a slowly varying envelope term $\boldsymbol{e}(\boldsymbol{r})$. Accordingly, in our RWM, we expand the electric field as $\boldsymbol{E}(\boldsymbol{r})=\sum_{j=1}^{N}u_j\boldsymbol{M}_j(\boldsymbol{r})$ and adopt a novel multi-scale basis function  $\boldsymbol{M}_j(\boldsymbol{r}) $
 containing  fast oscillation  term  $e^{-i \phi(\boldsymbol{r})}$ and slowingly varying term $\boldsymbol{e}(\boldsymbol{r})$ as given by
\begin{equation}
\boldsymbol{M}_j(\boldsymbol{r})  =\boldsymbol{N}_j(\boldsymbol{r})e^{-i\phi(\boldsymbol{r})},\label{multiBF}
\end{equation}
where polynomial function $\boldsymbol{N}_j(\boldsymbol{r})$ denotes the $j\text{-th}$ basis function, and $\phi(\boldsymbol{r})$ is the known propagation phase distribution calculated from FIM in geometry-optics limitation. Following the standard procedure of FEM and choosing the expansion  and test functions as $\boldsymbol{M}_j(\boldsymbol{r})$ and $\boldsymbol{M}_i(\boldsymbol{r})^*$ , respectively, we can obtain the weak form in practical FEM implementation that  corresponds to the vector wave equation in \Eq{wave_eq}, 
\begin{equation}\label{weak_form}
\begin{split}
\int_{\Omega}\left[\nabla\times \boldsymbol{M}_i^*(\boldsymbol{r}) \cdot \bar{\bm {\mu}}_r^{-1} (\bm r)\nabla \times \boldsymbol{M}_j(\boldsymbol{r}) -k_0^2 \bar{\bm {\epsilon}}_r (\bm r) \boldsymbol{M}_i^*(\boldsymbol{r})\cdot \boldsymbol{M}_j(\boldsymbol{r})\right] dV \\
+\int_{{\Gamma}} \boldsymbol{M}_i(\boldsymbol{r})^* \cdot  \bar{\bm {\mu}}_r^{-1} (\bm r) \boldsymbol{n} \times \nabla\times \boldsymbol{M}_j(\boldsymbol{r}) dS=0,
\end{split}
\end{equation}
where $\Gamma$ denotes the boundary that encloses the domain $\Omega$, and $\bar n$ denotes the outward unit normal vector to the boundary of the modeling domain. In Eq.~(\ref{weak_form}), on purpose, we choose the expansion and test functions being a complex conjugate pair to cancel the rapidly varying  factor as much as possible, which are given as follows
\begin{equation}\label{weak_form2}
\begin{split}
\int_{\Omega}\left[ \left( 
\nabla \times \boldsymbol{N}_i(\boldsymbol{r}) + i \nabla \phi(\boldsymbol{r}) \times \boldsymbol{N}_i(\boldsymbol{r}) 
\right) \cdot \bar{\bm {\mu}}_r^{-1} (\bm r) \left(
\nabla \times \boldsymbol{N}_j(\boldsymbol{r}) - i \nabla \phi(\boldsymbol{r}) \times \boldsymbol{N}_j(\boldsymbol{r}) 
\right) \right] dV \\
-\int_{\Omega} \left( k_0^2 \bar{\bm {\epsilon}}_r (\bm r) \boldsymbol{N}_i(\boldsymbol{r})\cdot \boldsymbol{N}_j(\boldsymbol{r})\right) dV +\int_{{\Gamma}} \left(\boldsymbol{N}_i(\boldsymbol{r}) \cdot \bar{\bm {\mu}}_r^{-1} (\bm r) \boldsymbol{n} \times (\nabla - i \nabla \phi(\boldsymbol{r})) \times \boldsymbol{N}_j(\boldsymbol{r})\right) dS=0.
\end{split}
\end{equation}
Notably, the validity of numerical implementation of \Eq{weak_form2} relies on the self-adjointness of the wave-equation operator $L$ under the complex inner product,
\beq\label{selfadjont}
\left( {\bm F,L\bm E} \right) = \left( {L\bm F,\bm E} \right), 
\eeq 
where the wave-equation operator $L = \nabla\times \bar{\bm {\mu}}_r^{-1} (\bm r) \nabla\times -k_{0}^2\bar{\bm {\epsilon}}_r (\bm r)$. See more details in previous works \cite{xiong2021finite,friedman1990principles,chen2019generalized,xiong2017classification}. In this regard, the selected complex inner product in the weak form as well as the self-adjointness of $L$  given in  \Eq{selfadjont} together impose constraints on the material tensor  $\bar{\bm {\epsilon}}_r (\bm r)$ and $\bar{\bm {\mu}}_r (\bm r)$ , i.e., $\bar{\bm {\epsilon}}_r^T = \bar{\bm {\epsilon}}_r^*$ and $\bar{\bm {\mu}}_r^T =\bar{\bm {\mu}}_r^*$, implying that the materials are lossless.

As implied in \Eq{weak_form2}, the practical implementation of ray-wave modeling consists of major three ingredients:   (1) calculating the phase-propagating factor, (2) plugging the phase factor into  \Eq{weak_form2} for pretreatment of numerical integration and boundary conditions and (3) FEM implementation of \Eq{weak_form2}, which can be summarized into the following 5 steps illustrated in \Fig{fig1}(c),
\begin{enumerate}
  \item Physical models are established for various scenarios, and two meshes with varying sparsity levels are generated for FIM and FEM computations;
  \item  Solve the optical path length using FIM on FIM mesh, obtain the phase function from the optical path length, map the phase function on FIM mesh to FEM mesh using interpolation algorithms; 
  \item  Combine the envelope function with the phase function, construct multi-scale basis functions;  
  \item  Assemble the finite element matrix based on basis functions and specific boundary conditions; 
  \item  Solve the finite element matrix to obtain the electric field of the physical model. 
\end{enumerate}

To ensure self-consistency, in the next subsection, we will briefly discuss  the numerical procedures of calculating propagation phases $\phi$ in geometry-optics approximation using FIM, as well as the technical details of adapting propagating phase numerically into FEM. As a side remark, one notes that both FIM and FEM crucially rely on the mesh, which are actually two independent set of meshes in our ray-wave method. However, we do use the mesh data of FEM to interpolate the propagating phase    $\phi(\boldsymbol{r})$ calculated from FIM to preserve the numerical accuracy in solving  \Eq{weak_form2}. Lastly, due to the implementation of parallel solution strategies, the time required for FIM of computing the phase factor can be negligible compared to the solution time of FEM.

\subsection{Practical Implementation}

\subsubsection{FIM for solving the propagation phase in geometry-optics approximation}
We now implement the specific process of applying FIM to compute the phase $\phi$ based on FIM mesh as outlined in the workflow in \Fig{fig1}(c). This involves employing the Godunov upwind difference scheme\cite{sethian1999fast} to numerically solve the eikonal equation within FIM mesh,
\begin{subequations}
\begin{align}
\left\vert\nabla s(r) \right\vert & -n(r)=0,\label{eik_eq}\\
\text{max}(D^{-x}u,-D^{+x}u,0)^2 & +\text{max}(D^{-y}u,-D^{+y}u,0)^2=n(x)^2,\label{godunov}
\end{align}
\end{subequations}
where $s(r)$ represents the optical path length from the source and $n(r)$ describes the distribution of the refractive index. \Eq{godunov} provides the discretized form of the eikonal equation on a 2D Cartesian grid. Here, $u$ is the discrete approximation of phase $\phi$ at a specific node. The operations $D^{-p}$ and $D^{+p}$ denote the forward and backward difference operators, respectively, along the axis $p\in \{x,y\}$.

FIM is aimed to solve the eikonal equation to obtain the optical path length at the arrival point of the wavefront, wherein the optical path is converted to optical phase for the construction of multi-scale basis function. One notices that the phase obtained in FIM extends from 0 to positive infinity to preserve the  phase gradient  ($\nabla \phi(\boldsymbol{r})$) continuity, since  the  phase within the range of  $-\pi$ to $\pi$ in \Eq{weak_form2} is hardly solvable. In contrast to computing optical path using ray tracing algorithm,  we calculate the optical path through the wavefront in FIM, which to a certain extend incorporates the diffraction characteristics of light wave according to Huygens–Fresnel principle, thus enabling the calculation of phase information of the whole space.
 
Notably, FIM is unable to handle  the scenario where  interference occurs, i.e., more than two sets of wave-fronts intersect.  Specifically,  at points where light rays cross, the  solution of the eikonal equation only records the phase of the first arriving wave, thus  wavefront information of the rest waves are lost. Inspired by the bidirectional beam envelope method, the inclusion of multiple wave-fronts in RWM is possible but beyond the scope of the current paper, which will be studied in our future work.

\subsubsection{FEM for solving slowly varying envelope}
Provided the phase distribution $\phi(\boldsymbol{r})$ from FIM is known, we subsequently construct multi-scale basis function $\boldsymbol{M}_j(\boldsymbol{r})$ in conjunction with polynomial basis functions $\boldsymbol{N}_j(\boldsymbol{r})$ as depicted in \Fig{fig1}(c). These multi-scale basis functions are applied to assemble the finite element matrix in FEM to solve electric fields. Next, we will illustrate this process through a concrete example,  the computation domain of which is ${\Omega}$  surrounded by the closed surface  $\Gamma_1$ and $\Gamma_2$. Without loss the generality, the external  plane wave  excitation  $\boldsymbol{E}^{i} e^{-ik_0 \boldsymbol{k}^i\cdot \boldsymbol{r}}$ enters  FEM algorithm  through boundary condition  at $\Gamma_1$ boundary, as given by 
\begin{equation}
\label{eqBC1}
\begin{split}
    \boldsymbol{n} \times \nabla \times \boldsymbol{E}(\boldsymbol{r})-&i k_0 \boldsymbol{n} \times \boldsymbol{E}(\boldsymbol{r}) \times \boldsymbol{n}=\\&\boldsymbol{n} \times (\nabla \times \boldsymbol{E}^{i}(\boldsymbol{r}) )e^{-ik_0 \boldsymbol{k}^i\cdot \boldsymbol{r}}
    -ik_0 \boldsymbol{n} \times (\boldsymbol{E}^{i}(\boldsymbol{r}) \times(\boldsymbol{n}-\boldsymbol{k}^i ))e^{-ik_0 \boldsymbol{k}^i\cdot \boldsymbol{r}},
\end{split}
\end{equation}
where $\boldsymbol{k}^i$ is the incident direction, and $\boldsymbol{n}$ is the unit outward normal vector of the boundary, $ \boldsymbol{E}(\boldsymbol{r})$ is the dependent variable that we inteed to solve. The scattering light exits through the remaining boundary $\Gamma_2$, the boundary condtion of which reads, 
\begin{equation}
\label{eqBC2}
    \boldsymbol{n} \times \nabla \times \boldsymbol{E}(\boldsymbol{r})-i k_0 \boldsymbol{n} \times \boldsymbol{E}(\boldsymbol{r}) \times \boldsymbol{n}=0.
\end{equation}
By applying the standard FEM  procedure to Eq.(\ref{wave_eq}) in junction with boundary condition given by  Eq.(\ref{eqBC1},\ref{eqBC2}),  we derive the linear equation
\begin{equation}
\label{rwm_matrix}
    Ax=b.
\end{equation}
In this formulation, $A$ donates the stiffness matrix, with the element in the $i$-th row and $j$-th column given by:
\begin{equation}
\label{matrixA}
\begin{split}
A_{ij}=\int_{\Omega}\left[ \left( 
\nabla \times \boldsymbol{N}_i(\boldsymbol{r}) + i \nabla \phi(\boldsymbol{r}) \times \boldsymbol{N}_i(\boldsymbol{r}) 
\right) \cdot \bar{\bm {\mu}}_r^{-1} (\bm r) \left(
\nabla \times \boldsymbol{N}_j(\boldsymbol{r}) - i \nabla \phi(\boldsymbol{r}) \times \boldsymbol{N}_j(\boldsymbol{r}) 
\right) \right] dV \\
-\int_{\Omega} \left( k_0^2 \bar{\bm {\epsilon}}_r (\bm r) \boldsymbol{N}_i(\boldsymbol{r})\cdot \boldsymbol{N}_j(\boldsymbol{r})\right) dV +\int_{{\Gamma}_1+{\Gamma}_2} \left( \boldsymbol{N}_i(\boldsymbol{r}) \cdot i k_0  \boldsymbol{n} \times \boldsymbol{N}_j(\boldsymbol{r}) \times \boldsymbol{n} \right) dS.
\end{split}
\end{equation}
The vector $b$ represents the load vector, with the element in the $i$-th row given by: 
\begin{equation}
\label{matrixb}
\begin{split}
b_i=\int_{\Gamma_1} \left[ \boldsymbol{N}_i(\boldsymbol{r})e^{i\phi(\boldsymbol{r})} \cdot ik_0 \boldsymbol{n} \times \left(\boldsymbol{E}^{i}(\boldsymbol{r}) \times (\boldsymbol{n}-\boldsymbol{k}^i)\right)e^{-ik_0\boldsymbol{k}^i\cdot \boldsymbol{r}}\right]dS \\
-\int_{\Gamma_1} \left[\boldsymbol{N}_i(\boldsymbol{r})e^{i\phi(\boldsymbol{r})} \cdot \boldsymbol{n} \times \left(\nabla \times \boldsymbol{E}^{i}(\boldsymbol{r}) \right)e^{-ik_0\boldsymbol{k}^i\cdot \boldsymbol{r}} \right]dS.
\end{split}
\end{equation}
Due to the multi-scale nature of computation problems in RWM, where the computational domain significantly exceeds the working wavelength, we choose the polynomial basis function $\boldsymbol{N}_j(\boldsymbol{r})$ to be at least quadratic. To ensure accuracy of numerical integration, the integration of matrix elements must achieve at least fourth-order precision. By solving Eq. (\ref{rwm_matrix}), we can determine the electric field distribution within the computational domain. Notably, because the multi-scale basis function $\boldsymbol{M}_j(\boldsymbol{r})$ is complex, the stiffness matrix $A$ is non-symmetric but Hermitian, thus necessitating an appropriate solution method.

\subsection{Error Analysis}
\begin{figure}[h]
    \centering
    \includegraphics[width=1\linewidth]{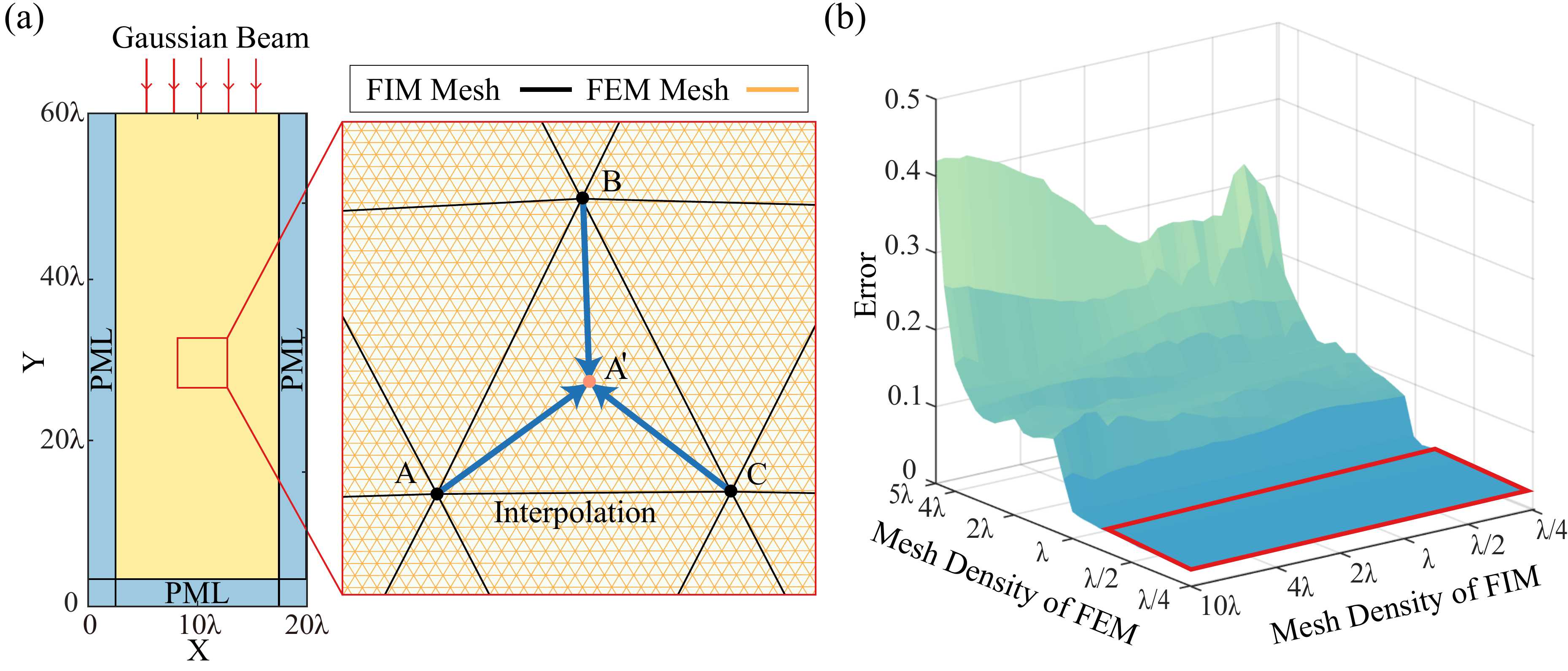}
    \caption{The results of error analysis for RWM. (a) Model structure. (b) Surface plot of Relative error  as function of FIM mesh density and FEM mesh density.}
    \label{fig2}
\end{figure}
We proceed to discuss the numerical performance of RWM, i.e., the relative error and its dependence as function of mesh density, which in this context refers to the edge length of the largest triangular element within the mesh. As sketched in \Fig{fig2}(a), we consider a Gaussian beam (beam waist is $2\lambda$, i.e., twice of the operation wavelength $\lambda$) propagating in grading refractive index medium, with distribution given by  $n=\sqrt{1+0.01x}$. The inset of \Fig{fig2}(a) shows the two sets of meshes, i.e., FIM mesh and FEM mesh as colored by solid black lines and solid orange lines respectively, located in the middle  of \Fig{fig2}(a) as indicated by the red square. Though FIM mesh is over 10 times coarser than FEM mesh, FIM can still obtain phase values rather accurately by interpolation on FEM mesh points from the values on FIM mesh points. 

\Figure{fig2}.(b) displays the relative error versus mesh densities, wherein the  relative error is  defined as follows,
\beq\label{error}
\text{error}=\sqrt{\frac{\sum_{i=1}^{n}\lVert E-E_{0} \rVert^2}{\sum_{i=1}^{n}\lVert E_{0} \rVert^2}},
\eeq
where $E_0$ is approximately the ground true value of the electric field  obtained  from fullwave FEM commercial software package MULTI-PHYSICS COMSOL, and $E$ is the electric field calculated using RWM. Evident from \Fig{fig2}.(b), as  the mesh density of FIM remains unchanged, the overall error of RWM decreases as the mesh density of FEM increases. While the error of RWM hardly changes against the variation of FIM mesh density for constant value of FEM mesh desity. This is because the phase distribution on the sparse FIM mesh can be mapped to a denser FEM mesh via barycentric interpolation for triangles. The application of the Godunov upwind difference scheme ensures accuracy in solving the eikonal equation on the sparse FIM mesh, allowing most of the phase information to be retained with minimal error.

Due to the intrinsic error of FIM itself at the computation nodes,  the numerical error continuously accumulates from the initial point to the surrounding area as the wavefront advances. Nevertheless, as can be observed from \Fig{fig2}(b), within the range marked by the red box, the RWM's errors remain consistently low. In particular, when the mesh densities of both FEM and FIM are $\lambda$, the error is only $0.0340$. In contrast, when the mesh density is set to $\lambda/5$, the error of FEM's results is $0.0264$. These data points imply that the aforementioned numerical error from FIM can be neglected. In this regard, we choose FIM mesh density to be $10\lambda$ to rapidly obtain phase values, and FEM mesh density  to be the value  ranging  from $\lambda$ to $\lambda/2$ in practical implementation of RWM. Next, we benchmark the computational time and the memory consumption of RWM with aforementioned  consideration of mesh sizes against the results calculated from commercial software package (Multiphysics COMSOL) through two concrete examples.

\section{Two numerical examples}
\subsection{Projection Objective Lens}

\begin{figure}[h]
    \centering
    \includegraphics[width=0.95\linewidth]{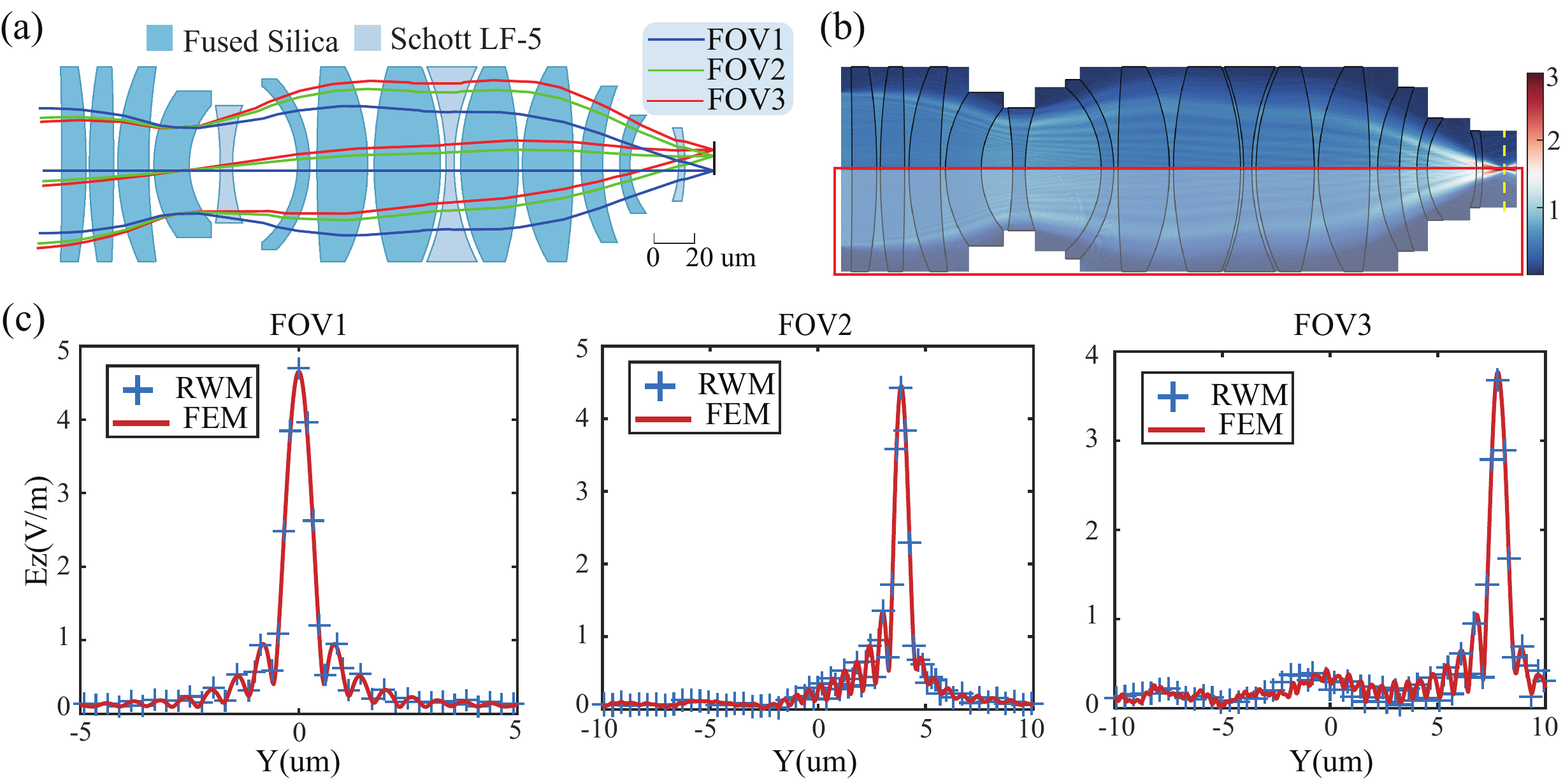}
    \caption{Simulation results of the projection objective lens.(a) Model structure. (b) Electric field simulation results of FEM and RWM with FOV1. (c) Comparison of electric fields simulated by RWM and FEM at the image plane under three different fields of view.}
    \label{fig3}
\end{figure}

The first example is the key component in  optical lithography machines, i.e., the projection objective lens, which usually consists of  multiple complex lens systems. Attaining high-performance in this aspect is critical for the evolution of lithography technology, directly influencing the precision, speed, and cost efficiency of chip fabrication processes\cite{Ulrich2002,stulen1999extreme}. Here, we provide an example to demonstrate the advantages of RWM in simulating projection objective lens. The example is based on a patented system\cite{shafer1988lens}, with its size proportionally reduced to fit the simulation range of RWM. \Figure{fig3}(a) illustrates the structure, which consists of 14 lenses and has overall dimensions of 80 µm by 300 µm. The incident light is a cylindrical wave with a wavelength of 405 nm, and simulations are conducted for three selected fields of view (FOVs). For the first FOV, the light source is aligned with the optical axis, while for the second and third FOV, light sources are positioned 20 µm and 40 µm away from the optical axis, respectively. In practical cases, anti-reflective coatings are applied to the lens surfaces to reduce stray light from Fresnel reflections. In RWM simulation, it is necessary to add transition boundary conditions on the lens surface to simulate the effect of the anti-reflective coating, as Fresnel reflections from the discontinuity in the refractive index can lead to interference fringes. Such fringes disrupt the slow variation of the envelope function, causing an artificial destructive interference with the incident beam. This process effectively eliminates the influence of interference fringes and enhancing the simulation accuracy.

\Figure{fig3}(b)  presents a comparison of the electric field simulation results obtained using RWM and the standard FEM for FOV 1. The upper part of the figure displays the results from RWM, while the lower part (within the red box) shows the results from FEM. \Figure{fig3}(c) illustrates the comparison of the electric field intensity between FEM simulation and RWM simulation at the focal plane,  specifically along the yellow dashed line in \Fig{fig3}(b), under three FOVs. The primary and secondary maxima of RWM closely align with the standard FEM results across the three FOV.

Table.\ref{po_data} presents the calculation information and errors of two simulation methods. The error is calculated using the method detailed in \Eq{error}. The reference electric fields are obtained from standard FEM simulations with a mesh density of $\lambda/12$. In RWM simulation, FIM mesh density is $10\lambda$ and FEM mesh density is $\lambda/1.5$, whereas the standard FEM simulation use a mesh density of $\lambda/5$. For the mesh density of $\lambda/1.5$, the total number of elements in the model is 1,387,527, and the degrees of freedom (DOFs) is 2,800,448. Conversely, for a mesh density of $\lambda/5$, the model contains 15,693,318 elements and 31,470,915 DOFs. Clearly, compared to the standard FEM, RWM uses only one-tenth of the DOFs, which results in both the simulation time and the computational resources required for RWM are only one-tenth of those required for standard FEM, while maintaining a comparable computational error to that of FEM. Because a very sparse mesh is used to solve the phase, the computation time for FIM, added with interpolation time, is less than one second and can thus be considered negligible. Additionally, in the third FOV, the large incident wave angle prevents the transition boundary conditions from eliminating all reflected waves. Consequently, interference occurs on certain surfaces, producing stray light and causing significant errors in RWM. However, this does not affect the accuracy of the focal plane's electric field.

\begin{table}[htbp]
\small
    \centering
    
    \caption{\bf Calculation information of RWM and FEM for projection objective lens.}
    \begin{tabular}{c|c|c|c|c|c}
    \hline
    \multirow{2}*{Methods}
     & Degree of  & \multirow{2}*{View} 
     & Simulation  & Computational
     & Relative \\
    & Freedom & & Time & Resources &  Error
    \\\cline{2-5}
    \hline\hline
    
    \multirow{3}*{FEM}
        & \multirow{3}*{31470915}
        & FOV 1 & 1403.21s & 214.84GB & 0.0528  \\
        & & FOV 2 & 1404.11s & 213.67GB & 0.0557  \\
        & & FOV 3 & 1405.32s & 213.05GB & 0.0662   \\\hline 

    \multirow{3}*{RWM}
        & \multirow{3}*{2800448}
        & FOV 1 & 54.21s & 25.19GB & 0.0958  \\
        & & FOV 2 & 60.34s & 24.40GB & 0.1029  \\   
        & & FOV 3 & 51.90s & 23.77GB & 0.1361  \\\hline

    \end{tabular}
    \label{po_data}

\end{table}


Furthermore, our local server has a maximum memory capacity of $1548GB$, eliminating the need for external storage during computations. If memory is limited, the computation time for standard FEM will greatly increase, further  underscoring the advantages of RWM. In summary, RWM can greatly reduce the DOF required for computation while maintaining high accuracy, representing a significant improvement over standard FEM. Furthermore, it can support various forms and directions of incident waves. However, it may overlook the interference fields generated, potentially affecting accuracy.

\subsection{Gradient refractive index Lenses}
 
Gradient refractive index (GRIN) lenses, known for their exceptional broadband performance, high directivity, and energy transmission efficiency, are widely employed in various applications\cite{moore1980gradient}. However, the inherently complex refractive index distribution of GRIN lenses introduces significant challenges for conducting precise simulations. Here, we use Mikaelian lens\cite{mikaelian1980v}, self-focusing fiber and Luneburg lens\cite{kundtz2010extreme} as three examples to verify the effectiveness of RWM in simulating GRIN lenses. The refractive index distribution of these lenses can generally be expressed as:

\begin{subequations}
\begin{align}\label{grin_lens}
n_{1}(y) & =n_0 \text{sech}(2 \pi py/d)\\
n_{2}(y) & =n_0(1-Ay^2)\\
n_{3}(r) & =\frac{1}{f}\sqrt{1+f^2-(\frac{r}{R})^2}
\end{align}
\end{subequations}
where $n_{1}$, $n_{2}$, and $n_{3}$ represent the refractive index distribution of Mikaelian lens, self-focusing fiber and Luneburg lens, respectively. $d$ is the thickness of Mikaelian lens, $n_0$ is the central refractive index along the optical axis, $p$ and $A$ are constants that determine the rate of change of the refractive index. $R$ denotes the radius of the Luneburg lens, and $r$ represents the radial coordinate originating from the center of the lens. The dimensionless parameter $f$ determines whether the focal point is located inside or outside the lens, i.e., the focal point is on the surface of the lens when $f = 1$.

\begin{figure}[h]
    \centering
    \includegraphics[width=0.95\linewidth]{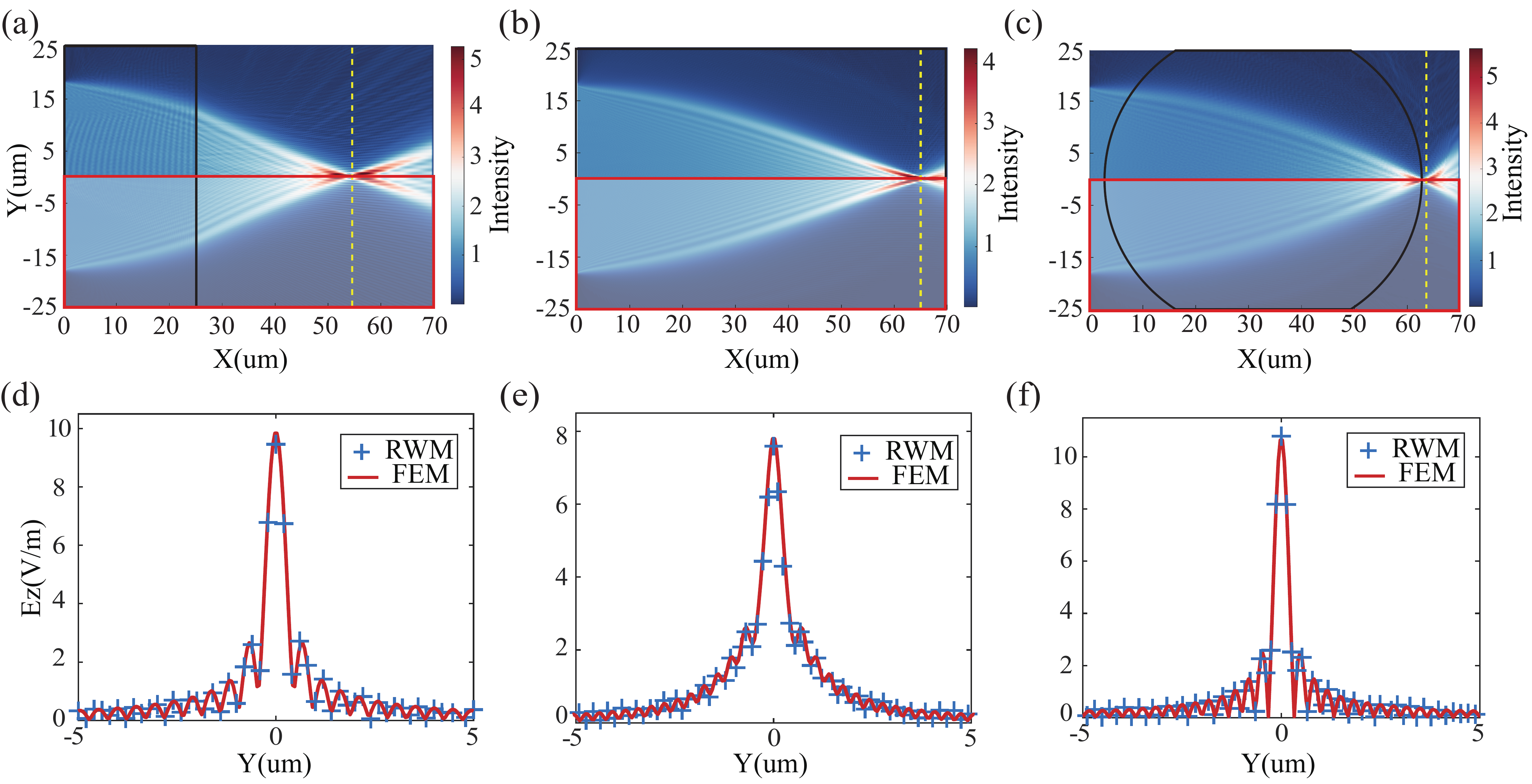}
    \caption{Simulation results of the gradient refractive index lenses. (a-c) Simulation results of Mikaelian lens, self-focusing fiber and Luneburg lens by RWM and FEM, respectively. (d-f) Comparison of electric fields simulated by RWM and FEM at the focal plane of different lenses.}
    \label{grin_fig}
\end{figure}

Figures~\ref{grin_fig}(a-c) display the electric field intensity distributions obtained from RWM and standard FEM for three types of lenses. The entire simulation domain measures 50 um by 70 um, with a plane wave of 400 nm wavelength incident horizontally. The upper part of the figure presents the simulation results from RWM with FIM mesh density of $2\lambda$ and FEM mesh density of $\lambda/1.5$. In contrast, the bottom half in red box displays the simulation results of standard FEM, utilizing a mesh density of $\lambda/5$. The region delineated by the solid black line represents the area of the lens, while the yellow dashed line indicates the focal plane. Figures~\ref{grin_fig}(d-f) display the comparison of electric fields simulated by RWM and standard FEM at the focal plane. The simulation results of RWM, whether for the main peak or the secondary peak, closely match those of FEM. This demonstrates that RWM is capable of achieving high accuracy simulation of the gradient refractive index system , even when the mesh density is $\lambda/1.5$.

\begin{table}[h]
    \centering
    \caption{\bf Calculation information of RWM and FEM for gradient refractive index lenses.}
    \begin{tabular}{*{4}{c|}c}
    \hline
    \multirow{2}*{Methods}
     &  \multirow{2}*{Lens Type}  &Degree of
     & Simulation  &  Relative \\
    & & Freedom & Time & Error
    \\\cline{2-5}
    \hline\hline
    \multirow{3}*{FEM}& Mikaelian Lens & 4284775 & 111.93s
    & 0.0740 \\
    & self-focusing fiber & 6865537 & 191.81s & 0.0785\\
    & Luneburg Lens & 5427785 & 126.46s  & 0.0266\\\hline
    
    \multirow{3}*{RWM}& Mikaelian Lens & 408797 & 7.73s
    & 0.1089 \\
    & self-focusing fiber & 583337 & 10.78s & 0.0970\\
    & Luneburg Lens & 512716 & 10.37s & 0.0919\\\hline
    \end{tabular}
    \label{lb_data}

\end{table}

Table.\ref{lb_data} presents simulation information. The reference electric fields are conducted from FEM simulation results with a mesh density of  $\lambda/15$. Comparing the simulation data between FEM and RWM, it becomes evident that with a reduction in DOFs to one-tenth, the time and computational memory required for RWM are also reduced to merely one-tenth of that needed by the standard FEM. Despite this reduction, the simulation error for RWM remains relatively small.

\section{Conclusion}
In conclusion, we introduce a  ray-wave combined numerical algorithm that efficiently and accurately performs simulations of multi-scale optical devices/systems. The phase distribution of electric field within the calculation domain is derived by solving the eikonal equation using FIM. A novel multi-scale basis function is then developed by combining polynomial basis functions. Utilizing this multi-scale basis function, FEM can accurately simulate the electric field in optical systems with fewer DOFs, thus enabling the simulation of large multi-scale optical systems. The proposed method performs integrated simulations without separating different devices within the system or involving complex data exchanges between algorithms. Numerical simulations of lens arrays and gradient refractive index lenses validate the accuracy and efficiency of the proposed algorithm. We believe that this method represents a significant advancement in optical system simulation, aiding researchers in more effectively designing and analyzing multi-scale optical devices/systems.


\section*{Funding}

National Natural Science Foundation of China (Grant No. 61405067), National Natural Science Foundation of Hubei Province (Grant No. 2024AFA016) and the Innovation Project of Optics Valley Laboratory .

\section*{Disclosures}

The authors declare that there is a potential conflict of interest regarding the work described in this manuscript. Yuntian Chen, Jingwei Wang and Fan Xiao have filed a patent application related to the method discussed in this paper. The application is currently under review. The details of the patent application are as follows: CN202410681076.9. This patent has not influenced the representation or interpretation of the reported research results.
\addcontentsline{toc}{chapter}{Bibliography}
\bibliography{main_paper}

\end{document}